\documentclass[aps,prd,groupedaddress,draft]{revtex4}
\usepackage{amsmath}
\usepackage{slashed}
\begin{document}


\voffset1.5cm

\title{Particle Production at High Energy and Large Transverse Momentum - "The Hybrid Formalism" Revisited.}
\author{ Tolga Altinoluk and Alex Kovner}
\affiliation{
 Physics Department, University of Connecticut, 2152 Hillside
Road, Storrs, CT 06269-3046, USA}
\date{\today}

\begin{abstract}
We revisit the "`hybrid formalism"' for particle production used recently to study saturation effects in single hadron multiplicities at forward rapidities at RHIC and LHC. We point out that at leading twist there is an extra contribution to the formulae used so far, which corresponds to particle production via inelastic scattering of the projectile partons on the target fields. This contribution is expected to be small due to kinematics at very forward rapidities/very high transverse momenta, but should be significant at high momenta and very high energies. This contribution is expected to be most affected by saturation effects and is therefore an interesting object of study in the context of possible onset of saturation at high energies.
\end{abstract}
\maketitle
\section{Introduction.}
In the last several years a large amount of RHIC data has been interpreted in terms of perturbative saturation physics\cite{saturation}, or the so called Color Glass Condensate (CGC)\cite{cgc}. Although this interpretation is philosophically very simple and appealing, alternative interpretations are also available and it is important to be able to differentiate between them. One would like to understand to what extent the data really unambiguously supports the idea of saturation. One of the problems we are faced with in this regard, is that the saturation based calculations, although in principle rooted in controlled perturbative approach to fundamental QCD physics, in practical implementations rely on phenomenological or semi - phenomenological ansatze and variety of shortcuts. As a result it is sometimes difficult to understand what features of the theoretical results are genuine and robust predictions of saturation, and which are model dependent transient features. The saturation based calculational techniques have advanced considerably during the last two-three years. In particular large part of next to leading corrections\cite{nexttoleading} is now taken into account in calculating the evolution of gluon density to high energy. This allowed for good fits to the HERA DIS data at low x\cite{guilherme}.
Still there is a lot of room for improvement, and in our view this improvement is absolutely essential in order for comparisons with data to have clear value. It is especially important at this time, since lately many aspects of the RHIC (and LHC) data have been analyzed in the framework of saturation physics\cite{footnote1}. 
These include the single inclusive particle production\cite{inclusive}, \cite{albacete} and the two hadron correlations at forward rapidity\cite{corr}, \cite{albacete1} in dA collisions, 
and more recently description of single particle spectra\cite{pp} and attempts to understand  \cite{ridgetheory} ridge in p-p collisions at LHC\cite{cms}. 
Although qualitatively the agreement of saturation based approaches with data is heartening, the calculations have still a long way to go before they can achieve the level of reliability comparable to that of, for example standard perturbative QCD approach.

The present paper is a modest contribution to improvement of one particular calculational approach addressing one particular piece of data - particle production in deuteron - gold collisions at forward rapidities. RHIC experiments observed strong suppression in the particle production in dA at forward rapidities. The "`state of the art"' saturation calculation of this effect appears in \cite{albacete}. Although the data is described quite well, there are some peculiarities to the results of \cite{albacete}. First, a very small $K$-factor is required to fit the overall magnitude of the production of neutral pions, while no $K$ - factor is required to fit the charged hadron multiplicity. Secondly and perhaps more worryingly, the suppression in the theoretical curves of \cite{albacete} when extended to LHC energies persists to extremely high transverse momenta, where one expects perturbation theory to be long applicable and $R_{dA}$ to be equal to one.  

The calculations of \cite{albacete} are based on the "`hybrid formalism"' of \cite{adrian}. In this approach the wave function of the projectile at large values of $x$ is calculated perturbatively, without soft approximation, while the scattering of the projectile partons on the target fields is treated in the eikonal approximation. The exact treatment of the projectile function is of course necessary to describe particle production at forward rapidity, since these partons cannot be in any way considered soft. It has another advantage over the soft approximation in the projectile wave function since it satisfies energy (longitudinal momentum) conservation for the incoming projectile. The energy conservation in the scattering process is still violated of course, since the recoil (and radiation) in the scattering event itself is not taken into account in the eikonal approximation. Although energy conservation must be very important at large values of $x_F$ and its effect has to be understood to make sure the treatment is consistent, we have nothing new to add to this point. 
In this paper we will revisit the derivation of inclusive particle production within the hybrid formalism {\it per se} relaxing only the collinear approximation made in \cite{adrian}. Our goal is to identify the terms which where omitted in \cite{adrian} but may nevertheless be important when the transverse momentum of produced particles is significantly higher than the saturation momentum of the target. As we will show, such terms, which do not correspond to collinear emission of the incoming projectile partons do indeed exist and contribute at leading twist. These terms have a simple physical interpretation and also have a simple form amenable to numerical implementations. It is out hope that including these terms in numerical calculations will improve the quality of the results.

The expression derived in \cite{adrian} and used in \cite{albacete} for particle production has a very intuitively appealing and simple form
\begin{equation}
\frac{dN}{d^2kd\eta}=\frac{1}{(2\pi)^2}\int_{x_F}^1 \frac{dz}{z^2} \Bigg[x_1f_g(x_1,Q^2)N_A(x_2,\frac{k}{z},b=0)D_{h/g}(z,Q)+\Sigma_qx_1f_q(x_1,Q^2)N_F(x_2,\frac{k}{z},b=0)D_{h/q}(z,Q)\Bigg]
\label{parton}
\end{equation}
where $N_{A(F)}(k)$ is the Fourier transform of the forward scattering amplitudes of the adjoint (fundamental) dipole.
It describes the process whereby incoming low $p_T$ partons scatter on the target independently of each other, acquiring large momentum $k_T$ in the process, and subsequently fragment into observed hadrons. This process is certainly the origin of large part of produced particles.

 However there is another physical mechanism which produces large $k_T$ particles in leading twist, whereby high $p_T$ particles preexisting in the wave function of the incoming projectile scatter with only a small momentum transfer from the target. The soft scattering is nevertheless enough to decohere the incoming partons from the rest of the wave function so that they materialize as on shell particles in the final state. The high $p_T$ partons in the projectile wave function arise due to DGLAP splitting of very forward partons.  The scattering process is essentially just the inelastic scattering of the forward projectile partons with emission of gluons (or quarks/antiquarks)\cite{lionia}. As we will show explicitly within the hybrid formalism, this mechanism of production is equally important as the one taken into account in eq.(\ref{parton}) when the saturation momentum of the target is small. When $Q_s$ is large, this contribution is somewhat suppressed, but may still be quantitatively quite large. Parametrically, while the contribution of eq.(\ref{parton}) is roughly  proportional to $\ln \frac{k_T}{ \Lambda_{QCD}}$, the additional inelastic scattering contribution scales like $\ln \frac{k_T}{ Q_s}$. It is thus only suppressed for $k_T\sim Q_s$ when $Q_s\ll\Lambda_{QCD}$, and even then the suppression is merely logarithmic. Given that for RHIC data $Q_s/\Lambda_{QCD}\sim 5$, it seems prudent to keep this contribution in numerical calculations.

It is quite clear that taking into account the inelastic mechanism must bring the calculation of particle production into agreement with the perturbative result at  large $p_T$. Thus we hope that including this contribution will bring $R_{dA}$ close to unity at reasonable values of $p_T$. It is also interesting to note that the final states of the inelastic scattering are quite different from those of the elastic one. The elastic piece is dominated by quarks in the final state, while the inelastic one contains comparable number of quarks and gluons. Since the fragmentation functions of quarks are very different, we expect it to affect the relative magnitude of neutral pion and charged hadron production and thus be relevant to the problem of a very small $K$ factor for neutral pion production encountered in \cite{albacete}.
Whether it helps or makes things worse remains to be seen. Naively one expects gluons to fragment predominantly into neutral mesons, and thus the problem of the $K$ factor may become even more acute, since the neutral to charged hadron ratio is likely to increase after including the inelastic contribution.

We also note that it is the inelastic term which is especially sensitive to the saturation effects. The wave function of the incoming hadron knows nothing about saturation by itself. The effects of saturation come entirely from the distributions of the target. The target fields are directly affected by saturation at momenta $k<Q_s$. The elastic scattering probes the large momentum component of target fields, equal to the final momentum of the produced parton. Thus as long as $p_T>Q_s$, this part of hadron production should be less affected by the saturation effects and one could expect that its dependence on energy and atomic number stems from perturbative physics. Any non leading twist scaling then presumably comes from effects of a possible "`nonperturbative"' initial condition propagated to higher momenta via perturbative evolution. The inelastic scattering contribution on the other hand probes the target fields at $k_T\ll p_T$ which includes the region $k_T< O(Q_s)$. It is this region of momenta which is strongly affected by target saturation effects. Thus if one neglects the inelastic contribution, one also severely limits ones options of studying effects of saturation. 

The structure of this paper is the following. In Sec.2 we revisit the derivation of particle production in the hybrid formalism. For simplicity of exposition we discuss in this section the pure glue theory.
In Sec. 3 we generalize our derivation to full QCD with quarks and antiquarks. We present a simple formula for the inelastic parton production contribution in terms of the  gluon and quark distribution functions. We conclude in Sec. 4 with discussion of our results.

\section{A prototypical calculation - gluon production}
In this section we derive the expression for gluon contribution to hadron production in the hybrid formalism. We will include the quark and antiquark contributions in the following section.
Our approach in the formal sense is similar to that of \cite{gluons}, although like in \cite{adrian} we are not approximating the gluon splitting function by its low $x$ limit. This will give us a possibility  to compare our results with the $k_T$ factorized formula which arises very simply in the approach of \cite{gluons}, to get some intuition from the simple
$k_T$ factorized expression and also to see the similarities and differences between the hybrid and the $k_T$ factorized results.
 
We consider a process where an energetic projectile scatters off a static target. The wave function of the incoming projectile is an eigenstate of the QCD Hamiltonian. When calculated in the perturbation theory it can be represented as
\begin{equation}
|\Psi\rangle_{in}=\Omega|v\rangle
\end{equation}
where $|v\rangle$ is the zeroth order wave function (an eigenfunction of the free Hamiltonian), and $\Omega$ is a unitary operator which diagonalizes the QCD Hamiltonian in perturbation theory
\begin{equation}
\Omega^\dagger H_{QCD}\Omega=H_{diag}
\end{equation}
The gluonic state immediately after scattering is
\begin{equation}
|\Psi\rangle_{out} =S|\Psi\rangle_{in}
\end{equation}
where $S$ is the eikonal scattering matrix for the projectile partons which propagate through the static target fields.

The number of produced gluons is then given by
\begin{equation}
\frac{dN}{d^2kd k^+}=\frac{1}{(2\pi)^3}\langle v|\Omega^\dagger S^\dagger \Omega a^\dagger(k,k^+)a(k,k^+)\Omega^\dagger S \Omega|v\rangle
\end{equation}

Our first goal is to find the operator $\Omega$. 
We start with the light-cone Hamiltonian of QCD:
\begin{equation}
H=\int_{k^+>0}\frac{dk^+}{2\pi}d^2z\bigg(\frac{1}{2}\Pi^-_a(k^+,z)\Pi^-_a(-k^+,z)+\frac{1}{4}G^{ij}_a(k^+,z)G^{ij}_a(-k^+,z)\bigg)
\end{equation}
where the electric and magnetic pieces have the form
\begin{eqnarray}
\Pi^-_a(x^-,x)&=&-\frac{1}{\partial^+}(D^i\partial^+A_i)^a(x^-,x)\nonumber \\
G^{ij}_a(x^-,x)&=&\partial_iA^a_j(x^-,x)-\partial_jA^a_i(x^-,x)-gf^{abc}A^b_i(x^-,x)A^c_j(x^-,x)
\end{eqnarray}
Our convention for the covariant derivative is
\begin{equation}
D^{ab}_i\Phi^b=\left(\partial_i\delta^{ab}-gf^{acb}A^c_i\right)\Phi^b
\end{equation}
We are working in the light cone gauge hence $A^+=0$ and as usual other light cone component of the vector potential $A^-$ is expressed via the solution of Maxwell's equations as $A^-=-\frac{1}{\partial^+}\partial_iA_i$. The transverse components of the vector potential $A^i$ which are the only dynamical degrees of freedom are expanded in the standard way in terms of the creation and annihilation operators
\begin{equation}
A^a_i(x^-,z)=\int_0^\infty \frac{dk^+}{2\pi}\frac{1}{\sqrt{2k^+}}\left\{a_i^a(k^+,z)e^{-ik^+x^-}+a_i^{a \dagger}(k^+,z)e^{ik^+x^-}\right\}
\end{equation}
where the creation and annihilation operators satisfy the canonical commutation relations
\begin{equation}
\left[a^a_i(k^+,x),a^{b \dagger}_j(p^+,y)\right]=2\pi\delta^{ab}_{ij}(k^+-p^+)\delta^2(x-y)\label{norm}
\end{equation}

We will calculate gluon production to the leading order in the coupling constant, and we therefore require to know the Hamiltonian only to first order in $g$. After 
some algebra we find
\begin{eqnarray}
H&=&H_0+H_1\\
H_0&=&\int_{k,k^+>0}\frac{k^2}{2k^+}a_j^{a\dagger}(k^+,k)a_j^a(k^+,k)\nonumber\\
H_1&=&-igf^{abc}\int_{k,p,k^+,p^+>0}\frac{1}{\sqrt{2k^+p^+(k^++p^+)}}\left\{-\left[\frac{p^+}{k^+}k_i-p_i\right]a_i^b(k^+,k)a^c_j(p^+,p)a_j^{a\dagger}(k^++p^+,k+p)\right. \nonumber \\ && \left. \qquad\ \ \ \ \ \ \ \ \ \ \ \ \ \ \ \ \ \ \ \ \ \ \ \  +\frac{p^+}{p^++k^+}k_ja_i^b(k^+,k)a_i^c(p^+,p)a_j^{a\dagger}(k^++p^+,k+p)\right\}+h.c.\nonumber
\end{eqnarray}
where the integration measure is understood as $\frac{dk^+}{2\pi}$ and $\frac{d^2k}{(2\pi)^2}$.

As $\Omega$ is a unitary operator, we define Hermitian operator $G$ by
\begin{equation}
\Omega=e^{-iG}=1-iG-\frac{1}{2}G^2+...
\end{equation}
The unitary operator $\Omega$ as discussed above is the operator that diagonalizes the Hamiltonian. To first order in the coupling constant the eigenvalues of the Hamiltonian are those of the free Hamiltonian $H_0$. To this order we have
\begin{equation}
\Omega^{\dagger} H\Omega=H-i[H,G]=H_0
\end{equation}
Thus the operator $G$ is determined from
\begin{equation}
i[H_0,G]=H_1
\end{equation}
This immediately gives
\begin{eqnarray}
G&=&-gf^{abc}\int_{k,p,k^+,p^+>0}\frac{1}{\sqrt{2k^+p^+(k^++p^+)}} \frac{1}{\omega_{p+k}-\omega_p-\omega_k}\\
&&\times
\left\{-\left[\frac{p^+}{k^+}k_i-p_i\right]a_i^b(k^+,k)a^c_j(p^+,p)a_j^{a\dagger}(k^++p^+,k+p)
\right. \nonumber \\ && \left. 
+\frac{p^+}{p^++k^+}k_ja_i^b(k^+,k)a_i^c(p^+,p)a_j^{a\dagger}(k^++p^+,k+p)\right\}+h.c.\nonumber
\end{eqnarray}
%
%
with
\begin{equation}
\omega(k)=\frac{k^2}{2k^+}
\end{equation} 

\subsection{Gluon Production}

The number of produced gluons to leading order in the coupling is given by
\begin{equation}
\frac{dN}{d^2kdk^+}=\frac{1}{(2\pi)^3}\langle v\vert\left[\hat{S}^{\dagger}G-G\hat{S}^{\dagger}\right]a_k^{a\dagger}(k^+,k)a_k^a(k^+,k)\left[G\hat{S}-\hat{S}G\right]\vert v\rangle
\end{equation}
Here the factor $\frac{1}{(2\pi)^3}$ is due to our normalization of the creation and annihilation operators eq.(\ref{norm}).
For simplicity of the calculation we will assume that the longitudinal momentum of the observed gluon is (at least) slightly smaller that the momentum of gluons in the state $|v\rangle$, although in fact our formulae will be valid in a more general case.

The calculation of the matrix element is straightforward. The $S$ matrix operator acts as a color rotation on all gluon creation and annihilation operators in coordinate space
\begin{equation}
\hat S^\dagger a^a_i(q^+,v)\hat S=S^{ab}(v)a^b_i(q^+,v)
\end{equation}
Since by assumption there are no gluons with longitudinal momentum $k^+$ in the state $|v\rangle$, one of the creation operators in the operator $G$ in the amplitude $\left[G\hat{S}-\hat{S}G\right]\vert v\rangle$ must be at momentum $k^+$ and is "contracted" with $a(k^+)$ in the observable. This then leaves us with (apart from the various factors of $S$, and omitting for simplicity transverse dependences) expectation value of the type
\begin{eqnarray}
\langle v|a^\dagger(p^++k^+)a(p^+)a^\dagger(q^+)a(q^++k^+)|v\rangle&=&\delta(p^+-q^+)\langle v|a^\dagger(p^++k^+)a(p^++k^+)|v\rangle\nonumber\\
&+&\langle v|a^\dagger(p^++k^+)a^\dagger(q^+)a(p^+)a(q^++k^+)|v\rangle
\label{contr}
\end{eqnarray}
The second term involves a two particle density in the state $|v\rangle$. It is suppressed in the leading twist "`partonic"' approximation. Since we keep to this approximation in the present paper, we neglect this term. We note that in the soft approximation, where the gluon production is given by the $k_T$ factorized expression of \cite{kovchegovtuchin} this term does indeed give a non vanishing contribution. We will make explicit connection with the soft approximation later.

Keeping only the first term in eq.(\ref{contr}) and reverting to coordinate space, where the $S$-matrix is diagonal we obtain
\begin{eqnarray}
\frac{dN}{ d^2kdk^+}&=&\frac{1}{ (2\pi)^3}\int e^{ik(z-\bar{z})}\langle v\vert\left[\hat{S}^{\dagger}G-G\hat{S}^{\dagger}\right]a_k^{a\dagger}(k^+,\bar{z})a_k^a(k^+,z)\left[G\hat{S}-\hat{S}G\right]\vert v\rangle \nonumber \\
&=&\frac{g^2}{(2\pi)^3}\frac{1}{ N_c^2-1}\int \frac{1}{k^+}e^{ik(z-\bar{z})+i\bar{p}v+i\bar{q}\bar{z}-i(\bar{p}+\bar{q})\bar{u}-ipv-iqz+i(p+q)u}tr\left\{\bigg[S^{\dagger}_{\bar{u}}T^{a}S_{\bar{u}}-S^{\dagger}_vT^{a}S_{\bar{z}}\bigg]\bigg[S^{\dagger}_{u}T^{a}S_u-S^{\dagger}_zT^{a}S_{{v}}\bigg]\right\}\nonumber \\&\times&
\frac{2}{(1-\xi)}\bigg[(1-\xi)^2+\xi^2+(1-\xi)^2\xi^2\bigg]\frac{\big[\xi\bar{p}_i-(1-\xi)\bar{q}_i\big]}{\big[\xi\bar{p}-(1-\xi)\bar{q}\big]^2}\frac{\big[\xi p_i-(1-\xi)q_i\big]}{\big[\xi p-(1-\xi)q\big]^2}
\langle a_j^{b \dagger}(\frac{k^+}{\xi},\bar{u})a_j^{b}(\frac{k^+}{\xi},u)\rangle
\end{eqnarray}
To arrive at this expression we assumed that the projectile state is color and rotationally invariant, so that
\begin{equation}
\langle a_i^{a \dagger}(p^+,\bar{u})a_j^{b}(p^+,u)\rangle=\frac{1}{2(N^2-1)}\delta^{ab}\delta_{ij}\langle a_k^{c \dagger}(p^+,\bar{u})a_k^{c}(p^+,u)\rangle
\end{equation}

Changing the integration variables
\begin{eqnarray}
\xi\bar{p}-(1-\xi)\bar q&=&\bar{\omega} \qquad \ \ \ \ \bar{p}+\bar{q}=\bar{\kappa} \nonumber \\
\xi p-(1-\xi)q&=&\omega \qquad \ \ \ \ p+q=\kappa
\end{eqnarray}
and integrating over $\omega$, $\bar{\omega}$, $\kappa$, $\bar{\kappa}$, $u$ and $\bar{u}$ one obtains
\begin{eqnarray}\label{finalobs}
\frac{dN}{d^2kdk^+}&=&\frac{\alpha_s}{ 2\pi^2}\frac{1}{(2\pi)^2}\frac{1}{N_c^2-1}\int_x^1\frac{d\xi}{\xi}\frac{1}{k^+}e^{ik(z-\bar{z})} \frac{2}{(1-\xi)}\bigg[(1-\xi)^2+\xi^2+(1-\xi)^2\xi^2\bigg]\frac{(v-\bar{z})_i}{(v-\bar{z})^2}\frac{(v-z)_i}{(v-z)^2}\nonumber \\
&\times&tr\left\{\bigg[S^{\dagger}((1-\xi)v+\xi\bar{z})T^{a}S((1-\xi)v+\xi\bar{z})-S^{\dagger}_vT^{a}S_{\bar{z}}\bigg]\bigg[S^{\dagger}((1-\xi)v+\xi z)T^{a}S((1-\xi)v+\xi z)-S^{\dagger}_zT^{a}S_{v}\bigg]\right\}
 \nonumber \\ &\times& \frac{k^+}{2\pi \xi}\langle a_j^{b \dagger}(\frac{k^+}{\xi},(1-\xi)v+\xi\bar{z})a_j^{b}(\frac{k^+}{\xi},(1-\xi)v+\xi z)\rangle
\end{eqnarray}
Throughout the rest of this paper we will continue using the notations $u$ and $\bar u$  to make notations less cumbersome, however it should be understood that they are not independent variables, but rather as shorthand for
\begin{equation}
u\equiv (1-\xi)v+\xi z; \ \ \ \ \ \bar u=(1-\xi)v+\xi\bar{z}
\end{equation}

To get some intuition about this expression we first consider the soft limit.

\subsection{The soft limit}
The soft limit corresponds to the situation when the longitudinal momentum of the observed gluon is much smaller than the momentum of the gluons in the valence state $|v\rangle$. Taking $\xi\rightarrow 0$ we obtain the this limit
\begin{eqnarray}\label{soft}
 \frac{dN}{ d^2k d k^+}&=&\frac{\alpha_s}{ \pi^2}\frac{1}{(2\pi)^2}\frac{1}{N_c^2-1}\int\frac{1}{k^+}e^{ik(z-\bar{z})}\frac{(v-\bar{z})_i}{(v-\bar{z})^2}\frac{(v-z)_i}{(v-z)^2}tr\left\{\bigg[S^{\dagger}_vT^{a}S_x-S^{\dagger}_vT^{a}S_{\bar{z}}\bigg] \bigg[S^{\dagger}_vT^{a}S_v-S^{\dagger}_zT^{a}S_{v}\bigg]\right\} \nonumber \\ &&\ \ \ \ \ \ \ \ \ \ \ \ \ \ \ \ \ \ \ \ \  \langle a_j^{\omega \dagger}(\frac{k^+}{\xi},v)a_j^{\omega}(\frac{k^+}{\xi},v)\rangle\\
&=&\frac{\alpha_s}{\pi^2}\frac{1}{(2\pi)^2}\frac{N_c}{N_c^2-1}\int \frac{1}{k^+}e^{ik(z-\bar{z})}\frac{(v-\bar{z})_i}{(v-\bar{z})^2}\frac{(z-v)_i}{(z-v)^2}tr\bigg\{1-S_v^{\dagger}S_z-S_{\bar{z}}^{\dagger}S_v+S_zS_{\bar{z}}^{\dagger}\bigg\}\langle a_j^{a\dagger}(\frac{k^+}{\xi}, v)a_j^{a }(\frac{k^+}{\xi},v)\rangle\nonumber
\end{eqnarray}
The $k_T$ factorized form reads (\cite{kovchegovtuchin} and also \cite{gluons},\cite{baier} corrected for typos) 
\begin{eqnarray}
 \frac{dN}{d^2k d\eta}&=&\frac{\alpha_s}{ \pi^2}\frac{1}{(2\pi)^2}\int e^{ik(z-\bar{z})}\left\{S_zS_{\bar{z}}^{\dagger}+S_vS_{\bar v}^{\dagger}-S_zS_{\bar v}^{\dagger}-S_vS_{\bar{z}}^{\dagger}\right\}^{ab}\frac{(z-v)_i}{(z-v)^2}\frac{(\bar{z}-\bar v)_i}{(\bar{z}
 -\bar v)^2}\langle \rho^a_v\rho^b_{\bar v}\rangle\nonumber\\
 &=&\frac{\alpha_s}{\pi^2}\frac{1}{(2\pi)^2}\int e^{ik(z-\bar{z})}\frac{1}{N_c^2-1}tr\left\{S_zS_{\bar{z}}^{\dagger}+S_vS_{\bar v}^{\dagger}-S_zS_{\bar v}^{\dagger}-S_vS_{\bar{z}}^{\dagger}\right\}\frac{(z-v)_i}{(z-v)^2}\frac{(\bar{z}-\bar v)_i}{(\bar{z}
 -\bar v)^2}\langle \rho^a_v\rho^a_{\bar v}\rangle
\label{softkt}
\end{eqnarray}
where the last equality follows from color neutrality of the hadronic state.
The color charge density operator here $\rho^a_v=\int \frac{dp^+}{ 2\pi} a^\dagger_i(p^+,v) T^aa_i(p^+,v)$. In the leading twist approximation the correlator of the charge density operators is local in the transverse space. Physically this is the case since in this (parton model) approximation there is only a small number of gluons in the hadron, and there are no correlations between different gluons. For a color singlet hadronic state we therefore have
\begin{equation}\langle \rho^a(v)\rho^a(\bar v)\rangle=\delta^2(v-\bar v)N_c\langle \int \frac{dp^+}{2\pi}a^{\dagger a}_i(p^+,v)a^a_i(p^+, v)\rangle
\end{equation}
Thus in the leading twist approximation eq.(\ref{soft}) is indeed equivalent to eq.(\ref{softkt}). 

It is customary to define the transverse momentum dependent gluon distribution in terms of the gluon distribution function $f_g(x, Q)$ as
\begin{eqnarray}
xf_g\left(x, Q=\frac{1}{ |u-v|}\right)\equiv \frac{ p^+}{2\pi}\int d^2 b \langle a^{a\dagger}_i(p^+,u)a^a_i(p^+,v\rangle)&=&\int d^2b \int \frac{d^2p}{\pi}e^{ip\cdot(u-v)}\phi(p,b;x)\nonumber\\
&\approx&\int d^2b \int_0^{\frac{1}{|u-v|^2}} dp^2\phi(p,b;x)
\end{eqnarray}
where $b=\frac{u+v}{ 2}$.
In the soft limit the color charge correlation function and the scattering amplitude are then expressed in terms of the projectile and the target distributions as
\begin{eqnarray}
\langle \rho^a(v)\rho^a(\bar v)\rangle& =&\frac{1}{ 8\pi \alpha_s}\int d^2pe^{ip\cdot(v-\bar v)}p^2\phi_P(p,b)\\
tr[1-S^\dagger(v)S(\bar v)]&=&2\pi\alpha_sN_c\int d^2pe^{ip\cdot(v-\bar v)}\frac{1}{p^2}\phi_T(p,b)
\end{eqnarray}
In terms of the transverse momentum distribution the single inclusive gluon spectrum in the soft limit has the familiar $k_t$ factorized form

\begin{eqnarray}
\frac{dN}{ d^2k d\eta}&=&S\frac{\alpha_sN_c}{ N_c^2-1}\int_l\bigg[\frac{1}{(l+k)^2}+\frac{1}{(l+k)^2}\frac{l^2}{k^2}+2\frac{1}{(l+k)^2}\frac{l\cdot k}{k^2}\bigg]\phi_T(l+k)\phi_P(l)\nonumber\\
&=&S\frac{\alpha_sN_c}{ N_c^2-1}\frac{1}{ k^2}\int_l \phi_T(l+k,Y-\eta)\phi_P(l,\eta)
\label{softlim}
\end{eqnarray}
where we have assumed translational invariance in the transverse plane. Here $S$ is the total transverse area of the collision and $Y$ denotes the total rapidity difference between the projectile and the target in the process.

In the limit of large momentum of the produced gluon $k\gg Q_s,\Lambda_{QCD}$ the momentum integral in eq.(\ref{softlim}) is dominated by two regions of momentum space.

In the first region, $l\ll k$ the dominant term is the first term in eq.(\ref{softlim}) (which corresponds to the first term in eq.(\ref{soft})).
In this kinematics the incoming projectile gluon has a small transverse momentum (in accordance with the simple parton model picture) and it acquires a large transverse momentum due to elastic scattering from the target field. We will refer to this contribution as elastic:
\begin{equation}
\Bigg[\frac{dN}{d^2k d\eta}\Bigg]_{elastic}=\frac{\alpha_sN_c}{ N_c^2-1}\frac{1}{k^2}\phi_T(k)\int_{l<Q\sim k}S\phi_P(l)\label{softelastic}
\end{equation}
The final states that correspond to this contribution have a single high $p_T$ gluon at forward rapidity. The balancing transverse momentum is carried by another gluon kicked out of the target by recoil, and it resides at a very different rapidity, close to the target.

There is however another contribution which is equally important at the leading twist. This comes from the momentum range $l=k+q$ with $q\ll k$. Changing variables from $l$ to $q$ the other contribution is clearly just the mirror image of eq.{\ref{softelastic}). For reasons explained below we will refer to it as inelastic contribution
\begin{equation}
\Bigg[\frac{dN}{d^2k d\eta}\Bigg]_{inelastic}=\frac{\alpha_sN_c}{ N_c^2-1}\frac{1}{k^2}\phi_P(k)\int_{q<Q\sim k}S\phi_T(q)\label{softinelastic}
\end{equation}
In this kinematics all terms in eq.(\ref{softlim}) are equally important.
This contribution corresponds to a projectile gluon coming in with large transverse momentum in the wave function  and subsequently scattering with small momentum transfer. The scattering practically does not add to gluons transverse momentum, but decoheres the gluon from the incoming hadronic wave function. One can naturally ask, how do high transverse momentum gluons find themselves in the projectile wave function. The answer is, that they are always there as "`unresolved"' components of the "`parton model"' gluons. A low $p_T$ gluon can split via a standard DGLAP evolution into a two gluon state with large relative transverse momentum. The "`parton model"' gluons are therefore not point like objects, but rather composites, which at first order in $\alpha_s$ contain an admixture of a two gluon state.

We stress that it is not the collinear part of the DGLAP kernel that is responsible for this structure. The collinear emission contributes to multiplication of low momentum gluons in the wave function ("`low"' momentum here technically means  momentum lower than that imparted by the scattering). The splitting however contains also an ultraviolet contribution, which produces gluon pairs with large relative transverse momentum, which sit close to each other in the impact parameter plane. It is the presence of this compact two gluon configuration that makes a projectile gluon behave as a composite object. When such a composite parton scatters inelastically off a soft target field, its different components can be put on shell, emerging as high $p_T$ partons in the final state. Correspondingly, the structure of the final state incidentally  is quite different from those that arise from the contribution eq.(\ref{softelastic}), as both high $p_T$ partons from the projectile wave function end up close to forward rapidity. The inelastic contribution therefore takes into account production of forward dijets with large $p_T$.

To demonstrate that this is indeed the correct interpretation of eqs.(\ref{softelastic},\ref{softinelastic}) consider scattering of a "`composite"' projectile gluon state
\begin{equation}
|in>=\Bigg[|v,a>+g T^a_{bc}f(v,z)|v,b;\ z,c>\Bigg]
\end{equation}
The amplitude of finding a two gluon component in the soft limit should be identified with the Wezsacker-Williams field $f(v,z)\propto \frac{(v-z)_i}{ (v-z)^2}$.
After propagation through the target
\begin{equation}
|out>=\Bigg[S^{ab}(v)|v,b>+g T^a_{bc}f(v,z)S^{bd}(v)S^{ce}(z)|v,d;\ z,e>\Bigg]
\end{equation}
Since the two gluon state admixture is perturbatively small, to leading order the elastically scattered state is simply
\begin{eqnarray}
|out>_{elastic}&=&S^{ab}(v)\Bigg[|v,b>+g T^b_{cd}f(v,z)(z)|v,c;\ z,d>\Bigg]\nonumber\\
&=&S^{ab}(v)|v,b>+g T^a_{bc}S^{bd}(v)S^{ce}(v)f(v,z)|v,c;\ z,e>
\end{eqnarray}
The amplitude of elastic scattering with momentum transfer $k$
is then given by $\int d^2v e^{ik\cdot v}[S^{ab}(v)-\delta^{ab}]$. We now square it, sum over the color indices of the final state, average over color in the initial state
and multiply by the number of gluons in the incoming wave function with small transverse momentum. This gives the number of gluons in the final state produced via elastic scattering of gluons in the projectile, and is precisely the contribution of eq.(\ref{softelastic}).

The inelastic component of the final state is
\begin{equation}
|out>_{inelastic}=|out>-|out>_{elastic}=g T^a_{bc}f(v,z)S^{bd}(v)\Bigg[S^{ce}(z)|v,d;\ z,e>-S^{ce}(v)|v,d;\ z,e>\Bigg]
\end{equation}
We assume that the scattering matrix $S(v)$ is a slowly varying function of the coordinate. Then the large transverse momentum of the gluon in the final state can only arise from a large relative momentum between the two gluons at $v$ and $z$, that is from the high momentum component of the amplitude $f(v,z)$. The amplitude of finding two gluons with large transverse momentum in the final state is given by
\begin{equation}
g T^a_{bc}\int d^2ze^{ik\cdot(z-v)}f(v,z)S^{bd}(v)\Bigg[S^{ce}(z)-S^{ce}(v)\Bigg]
\end{equation}
Squaring this, summing (averaging) over the final (initial) color indices, multiplying by the density of gluons at $v$ and integrating over the face of the projectile hadron we recover precisely the contribution of eq.(\ref{softinelastic}) in the form of the second line of eq.(\ref{soft}).
Thus we conclude indeed that the two leading twist contributions correspond to elastic and inelastic scattering of a projectile gluon on the target field.

At high $p_T$, both contributions are of the same order of magnitude. The probability to find a low $p_T$ gluon in the projectile is of order unity, but then the probability of scattering with large momentum transfer is small, of order $\alpha_s$. On the other hand, the probability to find a high $p_T$ parton in the incoming wave function is of order $\alpha_s$, however the probability of it scattering softly off the target is of order one. If one assumes that the projectile and target wave functions have standard perturbative behavior, $\phi=\frac{\mu^2}{ p^2}$, one finds
\begin{eqnarray}
\Bigg[\frac{dN}{d^2k d\eta}\Bigg]_{elastic}&=&\alpha_s\mu^2_P\mu^2_T\ln \frac{p^2}{ \Lambda^2_{QCD}}\nonumber\\
\Bigg[\frac{dN}{d^2k d\eta}\Bigg]_{inelastic}&=&\alpha_s\mu^2_P\mu^2_T\ln \frac{p^2}{Q_s^2}
\end{eqnarray}
where we have assumed that the perturbative behavior for the target kicks in at momenta above $Q_s$ . If the energy of the process is large enough so that the target distributions manifest extended geometric scaling\cite{geometric}, the $Q_s^2$ in the last equation should be substituted by a higher scale which marks the upper end of the geometric scaling window. At any rate, it is clear that at parametrically large transfer momentum the two contributions are comparable, and both must be kept. In the application we have in mind the transverse momentum is probably not much higher than $Q_s$ and so the inelastic contribution does not have a logarithmic enhancement. However the logarithm in the elastic contribution is also not very large, perhaps a factor of 3 or 4. Thus whether the inelastic contribution can be neglected or not is a numerical question, and one would be well advised not to through it away prematurely.

\subsection{Back to the hybrid formalism.}
We now return to eq.(\ref{finalobs}). Our aim is to identify the two contributions described above in this more general formula, and to write the leading twist result in as simple form as possible.

The elastic contribution as before corresponds to the region of the phase space where all the transverse momentum of the produced gluon originates from the momentum transfer from the target. This comes from the product of the two last terms in the brackets in eq.(\ref{finalobs}). Assuming that the momentum in the rest of the expression is small is equivalent to take $z=\bar z$ everywhere else apart from the scattering amplitude $S^\dagger(z)S(\bar z)$
\begin{eqnarray}\label{finalelastic}
\Bigg[\frac{dN}{ d^2k d\eta}\Bigg]_{elastic}&=&\frac{\alpha_sN_c}{2\pi^2}\frac{1}{(2\pi)^2}\int_{x_F}^1 \frac{d\xi}{\xi}\frac{2}{(1-\xi)}\bigg[(1-\xi)^2+\xi^2+(1-\xi)^2\xi^2\bigg]\int_z\frac{1}{ (v-z)^2}\frac{k^+}{2
\pi\xi}\langle a_j^{b \dagger}(\frac{k^+}{\xi},\bar u)a_j^{b}(\frac{k^+}{\xi},u)\rangle
\nonumber\\
&\times&\frac{1}{N_c^2-1}\int_{z-\bar z}e^{ik(z-\bar{z})}tr[S^\dagger(z)S(\bar z)]\\
&\simeq&\frac{\alpha_s}{\pi}\frac{1}{(2\pi)^2}\int_{p^2<Q^2}\frac {dp^2}{2p^2}\int_{x_F}^1\frac{d\xi}{\xi}P_{g/g}(\xi)x_Ff_g(\frac{x_F}{\xi},p^2)N_A(k)\nonumber
\end{eqnarray}
where
\begin{equation}
P_{g/g}=\frac{2N_c}{\xi(1-\xi)}\bigg[(1-\xi)^2+\xi^2+(1-\xi)^2\xi^2\bigg]
\end{equation}
is the standard gluon-gluon splitting function.
As explained in \cite{adrian} the first line is just the DGLAP contribution to the evolution of the gluon distribution. The "`parton model"' term is not present in our explicit formula because we have for simplicity assumed that the rapidity of the observed gluon is lower than that of any of the valence partons, and thus such gluons can be present in the wave function only as a result of evolution. This assumption is of course not necessary. Relaxing it restores the parton model contribution to the elastic gluon production. The elastic term then simply becomes the contribution discussed in \cite{adrian} (we do not include yet the fragmentation function contribution)
\begin{equation}\label{elasticadrian}
\Bigg[\frac{dN}{d^2k d\eta}\Bigg]_{elastic}= \frac{1}{(2\pi)^2}x_Ff_g(x_F,Q^2)N_A(k)
\end{equation}  
 with $N_A(k)=\frac{1}{N_c^2-1}\int d^2(z-\bar z)e^{ik\cdot(z-\bar z)}tr[S^\dagger(\bar z)S(z)]$; and the factorization scale $Q$ must be chosen so that it is of order but smaller than the external momentum $k$.

To extract the inelastic term in the leading twist approximation we note that it arises as the leading order expansion in powers of $|x-z|$ and $|x-\bar z|$, since the separation between the two gluons in the wave function must be much smaller than the typical variation scale of the scattering amplitude $S(x)$.
Referring back to eq.(\ref{finalobs}) we write
\begin{eqnarray}
S_{\bar u}&\simeq&S_v-\xi(v-\bar{z})_i\partial_iS_v; \ \ \ \ \ \ \ S_{u}\simeq S_v-\xi(v-z)_i\partial_iS_v\nonumber \\
S_{\bar{z}}&\simeq&S_v-(v-\bar{z})_j\partial_jS_v; \ \ \ \ \ \ \ S_z \simeq S_v-(v-z)_j\partial_jS_v
\end{eqnarray}
Then for the amplitude in eq.(\ref{finalobs}) we find
\begin{eqnarray}
&&tr\left\{\bigg[S_{\bar u}^{\dagger}T^{\sigma}S_{\bar u}-S_{v}^{\dagger}T^{\sigma}S_{\bar{z}}\bigg]\bigg[S_{u}^{\dagger}T^{\sigma}S_{u}-S_{z}^{\dagger}T^{\sigma}S_{v}\bigg]\right\}\nonumber \\ 
&=&N_c(v-\bar{z})_i(v-z)_j\left\{(1-\xi)^2+\xi^2\right\}tr\bigg[\partial_iS_v\partial_jS_v^{\dagger}\bigg]-2\xi(1-\xi)(v-\bar{z})_i(v-z)_jtr\bigg[T^{\sigma}S_v\partial_iS_v^{\dagger}T^{\sigma}S_v\partial_jS_v^{\dagger}\bigg] \nonumber\\
&=&\frac{N_c}{2}(v-\bar{z})\cdot(v-z)\left\{(1-\xi)^2+\xi^2\right\}tr\bigg[\partial_iS_v\partial_iS_v^{\dagger}\bigg]-\xi(1-\xi)(v-\bar{z})\cdot(v-z)tr\bigg[T^{\sigma}S_v\partial_iS_v^{\dagger}T^{\sigma}S_v\partial_iS_v^{\dagger}\bigg]
\end{eqnarray}
where the last line strictly speaking holds only after averaging over rotationally invariant target for which $\langle \partial_iS^\dagger\partial_jS\rangle=\frac{1}{2}\delta_{ij}\langle \partial_kS^\dagger\partial_kS\rangle$. 
One can further simplify this expression noting that
 $S_v\partial_iS_v^{\dagger}=\frac{1}{N_c}T^{a} tr[S_v\partial_iS^\dagger_vT^a]$ i.e. it is a "`pure gauge"' vector potential.
Then using simple color algebra we find
\begin{equation}
tr [S\partial_iS^\dagger T^aS\partial_iS^\dagger T^a]=-\frac{N_c}{2}tr[ \partial_i S^\dagger\partial_i S]
\end{equation}
So that we can write
\begin{eqnarray}
&&tr\left\{\bigg[S_{\bar u}^{\dagger}T^{\sigma}S_{\bar u}-S_{v}^{\dagger}T^{\sigma}S_{\bar{z}}\bigg]\bigg[S_{u}^{\dagger}T^{\sigma}S_{u}-S_{z}^{\dagger}T^{\sigma}S_{v}\bigg]\right\}\nonumber \\ 
&=&\frac{N_c}{ 2}\left\{1-\xi+\xi^2\right\}(v-\bar{z})\cdot(v-z)tr\bigg[\partial_iS_v\partial_iS_v^{\dagger}\bigg]\nonumber\\
&=&N_c^2\left\{1-\xi+\xi^2\right\}(v-\bar{z})\cdot(v-z)tr\bigg[\partial_iS_F(v)\partial_iS_F^{\dagger}(v)\bigg]
\end{eqnarray}
where we have given the final answer in terms of the fundamental representation matrices $S_F$.
It is clear from our derivation that in the above expressions the target field average should be understood as calculated with resolution $Q$, where just as in the elastic piece, the factorization scale $Q$ is of order of, but smaller than the transverse momentum of the observed gluon $Q<k$. 
\begin{equation}
tr\bigg[\partial_iS_F(v)\partial_iS_F^{\dagger}(v)\bigg]\rightarrow tr\bigg[\partial_iS_F(v)\partial_iS_F^{\dagger}(v)\bigg]_Q=N_c\int_{p^2<Q^2} p^2 N_F(p)
\end{equation}
The leading twist part of the inelastic contribution can therefore be written as
\begin{eqnarray}\label{og}
\Bigg[\frac{dN}{d^2k d\eta}\Bigg]_{inelastic}&=&\frac{\alpha_s}{2\pi^2}\frac{1}{(2\pi)^2}\frac{1}{ N_c^2-1}\int e^{ik(z-\bar{z})} \frac{(v-\bar{z})_i}{(v-\bar{z})^2}\frac{(v-z)_i}{(v-z)^2}(v-\bar{z})\cdot(v-z)tr\bigg[\partial_iS_F(v)\partial_iS_F^{\dagger}(v)\bigg]_Q
\nonumber\\
&\times&N_c^2\frac{2}{(1-\xi)}\bigg[(1-\xi)^2+\xi^2+(1-\xi)^2\xi^2\bigg]\left\{1-\xi+\xi^2\right\}\langle a_j^{b \dagger}(\frac{k^+}{\xi},\bar u)a_j^{b}(\frac{k^+}{\xi},u)\rangle
\end{eqnarray}

This expression can be rewritten in terms of the gluon distribution. To leading twist the dependence of the vacuum average $\langle a_j^{b \dagger}(\frac{k^+}{\xi},\bar u)a_j^{b}(\frac{k^+}{\xi},u)\rangle$ can be substituted by $\langle a_j^{b \dagger}(\frac{k^+}{\xi},v)a_j^{b}(\frac{k^+}{\xi},v)\rangle_Q$. Performing the Fourier transform we then obtain
\begin{equation}
\Bigg[\frac{dN}{ d^2k d\eta}\Bigg]_{inelastic}=\frac{\alpha_s}{\pi^2}\frac{N_c^2}{ N_c^2-1}\frac{1}{k^4}\int_{x_F}^1\frac{d\xi}{\xi}\left\{1-\xi+\xi^2\right\}P_{g/g}(\xi)x_Ff_g(\frac{x_F}{\xi}, Q)
\int_{p^2<Q^2}\frac{d^2p}{(2\pi)^2} p^2 N_F(p)
\end{equation}

Of course, in order to calculate the spectrum of produced hadrons we have to include the gluon fragmentation functions. We assume as always, that produced gluons fragment into hadrons independently Taking this into account our result for production is
\begin{eqnarray}\label{gluonf}
\frac{dN}{d^2k d\eta}&=&\int_{x_F}^1 \frac{dz}{z^2}D_{h/g}(z,Q)\Bigg[ x_1f_g(x_1,Q^2)N_A(x_2,\frac{k}{z},b=0)\\
&+&\frac{\alpha_s}{\pi^2}\frac{N_c^2}{ N_c^2-1}\frac{z^4}{k^4}\int_{x_1}^1\frac{d\xi}{\xi}\left\{1-\xi+\xi^2\right\}P_{g/g}(\xi)x_1f_g(\frac{x_1}{\xi}, Q) \int_{p^2<Q^2}\frac{d^2p}{(2\pi)^2} p^2 N_F(x_2,p,b=0)\Bigg]\nonumber
\end{eqnarray}
where
\begin{eqnarray}
N_A\left(k,b=\frac{\bar z+z}{2}\right)&=&\frac{1}{ N_c^2-1}\int d^2(z-\bar z)e^{ik\cdot(z-\bar z)}tr[S_A^\dagger(\bar z)S_A(z)];\\
N_F\left(k,b=\frac{\bar z+z}{2}\right)&=&\frac{1}{ N_c}\int d^2(z-\bar z)e^{ik\cdot(z-\bar z)}tr[S_F^\dagger(\bar z)S_F(z)]\nonumber
\end{eqnarray}
and the longitudinal momentum fractions (neglecting the hadron mass) are
\begin{equation}\label{xs}
x_F=\frac{ k}{\sqrt{s_{NN}}}e^\eta; \ \ \ \ x_1=\frac{x_F}{ z}; \ \ \ \ \ x_2=x_1e^{-2\eta}
\end{equation}

Eq.(\ref{gluonf}) is our result for hadron production in a toy theory that does not contain quarks. This is obviously not a good approximation to reality especially at forward rapidities, where the quark contribution must be the leading one. We now turn to generalizing the previous discussion by including the quark contribution.

\section{Taking care of quarks}
We now include the quark piece into the light cone QCD Hamiltonian
\begin{equation}
H_{q}=J^+\frac{1}{(P^+)^2}D_iE_i-\frac{1}{2}\Psi_+^{\dagger}\slashed {P_T}\frac{1}{P^+}\slashed {P_T}\Psi_+
\end{equation}
where 
\begin{equation}
J^+=-g\Psi_+^{\dagger}\tau^a\Psi_+
\end{equation}
and as usual
\begin{equation}
P^\mu=-iD^\mu; \ \ \ \ D_i=\partial_i+ig\tau^aA_i^a; \ \ \ \slashed {P_T}=\gamma^iP_i
\end{equation}
Here $\tau^a$ are the generators of $SU(N)$ group in the fundamental representation. 

The dynamical spinors $\Psi_+$ are defined in terms of creation and annihilation operators as\cite{lightcone}
\begin{equation}
\Psi_+^{\alpha}(x^-,x)=\frac{1}{(2\pi)^3}\int_{k^+>0}d^2kdk^+\bigg[w_sb_s^{\alpha}(k^+,k)e^{-ik^+x^-iqx}+w_{-s}d_s^{\dagger \alpha}(k^+,k)e^{ik^+x^-+ikx}\bigg]
\end{equation}
Here $b_s$ ($d_s$) are the quark (antiquark) annihilation operators,
and the basis spinors $w_s$ are
\begin{equation}
w_{\frac{1}{2}}=\begin{pmatrix} 1 \\ 0 \\ 0 \\0 \end{pmatrix} \ , \  w_{-\frac{1}{2}}=\begin{pmatrix} 0 \\ 0 \\ 0 \\1 \end{pmatrix}
\end{equation}
 The quark algebra is 
\begin{equation}
\{b_s(k^+,k),b_{s'}^\dagger({k'}^+,k')\}=\{d_s(k^+,k),d_{s'}^\dagger({k'}^+,k')\}=(2\pi)^3\delta(k^+-{k'}^+)\delta^2(k-k')\delta_{ss'}
\end{equation}
When expanded in terms of quark, anti-quark and gluon creation/annihilation operators $H_q$ can be written as
\begin{equation}
H_q=H_0+H_1
\end{equation}
where the free Hamiltonian $H_0$ is
\begin{equation}
H_0=\frac{1}{(2\pi)^3}\int_{k^+>0}d^2kdk^+\frac{k^2}{2k^+}\bigg[b_s^{\alpha\dagger}(k^+,k)b_s^{\alpha}(k^+,k)+d_s^{\alpha\dagger}(k^+,k)d_s^{\alpha}(k^+,k)\bigg]
\end{equation}
and $O(g)$ Hamiltonian $H_{1}$ is
\begin{eqnarray}
H_1&=&\frac{1}{2}g\tau^c_{\alpha\beta}\int_{p^+,k^+>0}\frac{1}{\sqrt{2q^+}}\bigg\{\frac{2p^++q^+}{p^+q^+}\bigg[\frac{p_i}{p^+}-\frac{q_i}{q^+}\bigg]\delta_{ij}\delta_{s',s}-i\epsilon^{ij}\sigma^3_{s',s}\frac{q^+}{p^++q^+}\bigg[\frac{p_i}{p^+}-\frac{q_i}{q^+}\bigg]\bigg\} \nonumber \\ 
&\times&\bigg[b_{s'}^{\dagger \alpha}(p^++q^+,p+q)b_s^{\beta}(p^+,p)a_j^c(q^+,q)+d_{s'}^{\alpha}(p^+,p)d_s^{\dagger \beta}(p^++q^+,p+q)a_j^c(q^+,q)\bigg] +h.c.\nonumber \\ 
&+&\frac{1}{2}g\tau^c_{\alpha \beta}\int_{p^+,k^+>0}\frac{1}{\sqrt{2(p^++q^+)}}\Bigg[\bigg\{\frac{q^+-p^+}{p^++q^+}\bigg[\frac{p_i}{p^+}-\frac{q_i}{q^+}\bigg]\delta_{ij}\delta_{-s',s}-i\epsilon^{ij}\sigma^3_{-s',s}\bigg[\frac{p_i}{p^+}-\frac{q_i}{q^+}\bigg]\bigg\}\nonumber\\
&\times& d_{s'}^{\alpha}(q^+,q)b_s^{\beta}(p^+,p)a_j^{\dagger c}(p^++q^+,p+q)+h.c.\Bigg]
\end{eqnarray}
Following the same steps as before, we find the operator that diagonalizes the Hamiltonian to first order 
\begin{equation}
\Omega=e^{iG}=1+iG-\frac{1}{2}G^2+...
\end{equation}
with
\begin{eqnarray}
G&=&\int_{p,q}\Bigg[\frac{1}{2\sqrt{2q^+}}\frac{1}{\omega_{p+q}-\omega_p-\omega_q}\bigg\{-i\frac{2p^++q^+}{p^++q^+}\bigg[\frac{p_i}{p^+}-\frac{q_i}{q^+}\bigg]\delta_{ij}\delta_{s',s}-\epsilon^{ij}\sigma^3_{s',s}\frac{q^+}{p^++q^+}\bigg[\frac{p_i}{p^+}-\frac{q_i}{q^+}\bigg]\bigg\}\nonumber\\
&\times&g\tau^c_{\alpha \beta}\bigg[b_{s'}^{\dagger \alpha}(p^++q^+,p+q)b_s^{\beta}(p^+,p)a_j^c(q^+,q)+d_{s'}^{\alpha}(p^+,p)d_s^{\dagger \beta}(p^++q^+,p+q)a_j^c(q^+,q)\bigg]\nonumber\\
&+&\frac{1}{2\sqrt{2(q^++p^+)}}\frac{1}{\omega_{p+q}-\omega_p-\omega_q}\bigg\{i\frac{q^+-p^+}{p^++q^+}\bigg[\frac{p_i}{p^+}-\frac{q_i}{q^+}\bigg]\delta_{ij}\delta_{-s',s}+\epsilon^{ij}\sigma^3_{-s',s}\bigg[\frac{p_i}{p^+}-\frac{q_i}{q^+}\bigg]\bigg\} \nonumber\\
&\times&g\tau^c_{\alpha \beta}d_{s'}^{\alpha}(q^+,q)b_s^{\beta}(p^+,p)a_j^{\dagger c}(p^++q^+,p+q)\nonumber\\
&+&h.c.\Bigg]
\end{eqnarray}

Again, following the same steps as before we calculate the number of produced gluons
\begin{eqnarray}
\frac{dN_g}{d^2kdk^+}&=&\frac{1}{(2\pi)^3}\int_{\bar z z}e^{ik(z-\bar{z})}\langle v|\big[\hat{S}^\dagger G-G\hat{S}^{\dagger}\big]a_i^{\dagger a}(k^+,\bar{z})a_i^a(k^+,z)\big[G\hat{S}-\hat{S}G\big]|v\rangle \\
&=&\frac{\alpha_s}{2\pi^2}\frac{1}{(2\pi)^2}\int_{\bar z z}e^{ik(z-\bar{z})}\frac{1}{k^+}(1+(1-\xi)^2)\frac{(v-\bar{z})_i}{(v-\bar{z})^2}\frac{(v-z)_i}{(v-z)^2}\nonumber \\ &\times&\left\{\bigg[S_F^{\dagger}\big(\bar u\big)\tau^a\bigg]_{\bar{\sigma}\kappa}-S^{ab}(\bar{z})\bigg[\tau^{b}S^{\dagger}_F(v)\bigg]_{\bar{\sigma }\kappa}\right\} 
\left\{\bigg[\tau^aS_F\big(u\big)\bigg]_{\kappa \sigma}-S^{\dagger c a}(z)\bigg[S_F(v)\tau^{c}\bigg]_{\kappa \sigma}\right\}\nonumber\\
&\times&\Bigg[\langle b_s^{\dagger \bar{\sigma}}\big(\frac{k^+}{\xi},\bar u\big)b_s^{\sigma}\big(\frac{k^+}{\xi},u\big)\rangle+\langle d_s^{\dagger {\sigma}}\big(\frac{k^+}{\xi},\bar u\big)d_s^{\bar\sigma}\big(\frac{k^+}{\xi},u\big)\rangle\Bigg]\nonumber
\end{eqnarray}
In the leading twist approximation there are, as before, contributions corresponding to elastic and inelastic partonic scattering. The elastic contribution can be written as 
\begin{eqnarray}
&&\left[\frac{dN_g}{d^2kd\eta}\right]_{elastic}=\frac{\alpha_s}{4\pi^2}\frac{1}{(2\pi)^2}\int_{x_F}^1\frac{d\xi}{\xi} e^{ik(z-\bar{z})}\frac{N_c^2-1}{N_c}\left(1+(1-\xi)^2\right)\frac{(v-\bar{z})_i}{(v-\bar{z})^2}\frac{(v-z)_i}{(v-z)^2}\frac{1}{N_c^2-1}tr\bigg[S_A^{\dagger}(z)S_A({\bar{z}})\bigg]\nonumber\\
&\times&\frac{k^+}{2\pi\xi}\bigg[\langle b_s^{\dagger \sigma}\big(\frac{k^+}{\xi},\bar u\big)b_s^{\sigma}\big(\frac{k^+}{\xi},u\big)\rangle+\langle d_s^{\dagger {\sigma}}\big(\frac{k^+}{\xi},\bar u\big)d_s^{\sigma}\big(\frac{k^+}{\xi},u\big)\rangle\bigg]\\
&\simeq&\frac{\alpha_s}{\pi}\frac{1}{(2\pi)^2}\int_{p^2<Q^2}\frac {dp}{p}\int_{x_F}^1\frac{d\xi}{\xi}P_{g/q}(\xi)x_F\Bigg[f_q(\frac{x_F}{\xi},p^2)+f_{\bar q}(\frac{x_F}{\xi},p^2)\Bigg]N_A(k)\nonumber
\end{eqnarray}
with
\begin{equation}
P_{g/g}=\frac{N_c^2-1}{2N_c}\frac{1+(1-\xi)^2}{\xi}
\end{equation}

This takes into account the DGLAP correction to gluon distribution due to quark(and antiquark)-gluon splitting. When written in terms of gluon distribution, this term is already contained in eq.(\ref{elasticadrian}).
The inelastic contribution is given by the region of phase space where the momentum transfer from the target is small. 
Hence one can expand the eikonal factors $S(u)$ around $S(v)$. The result is
\begin{eqnarray}
&&\left[\frac{dN_g}{d^2kd\eta}\right]_{inelastic}=\frac{\alpha_sN_c}{8\pi^2}\frac{1}{(2\pi)^2}\int_{x_F}^1 e^{ik(z-\bar{z})}\left(1+(1-\xi)^2\right)\frac{\big[(v-\bar{z})\cdot(v-z)\big]^2}{(v-\bar{z})^2(v-z)^2}\bigg[\bigg(1+(1-\xi)^2\bigg)-\frac{\xi^2}{N_c^2}\bigg] \nonumber\\
&\times&\frac{k^+}{2\pi\xi}\bigg[\langle b_s^{\dagger \sigma}\big(\frac{k^+}{\xi},\bar u\big)b_s^{\sigma}\big(\frac{k^+}{\xi},u\big)\rangle+\langle d_s^{\dagger {\sigma}}\big(\frac{k^+}{\xi},\bar u\big)d_s^{\sigma}\big(\frac{k^+}{\xi},u\big)\rangle\bigg]\frac{1}{N_c}tr\bigg[\partial_jS^{\dagger}_F(v)\partial_jS_F(v)\bigg]\\
&=&\frac{\alpha_s}{ 2\pi^2}\frac{N_c^2}{ N_c^2-1}\frac{1}{k^4}\int_{x_F}^1\frac{d\xi}{\xi}\bigg[\bigg(1+(1-\xi)^2\bigg)-\frac{\xi^2}{N_c^2}\bigg]P_{g/q}(\xi)x_F\Bigg[f_q(\frac{x_F}{\xi}, Q)+f_{\bar q}(\frac{x_F}{\xi}, Q)\Bigg]\int_{p^2<Q^2}\frac{d^2p}{(2\pi)^2} p^2 N_F(p)\nonumber
\end{eqnarray}

Similarly one can also calculate the number of produced quarks
\begin{eqnarray}
\frac{dN_q}{d^2kdk^+}&=&\frac{1}{(2\pi)^3}\int e^{ik(z-\bar{z})}\langle v|\big[\hat{S}G-G\hat{S}^{\dagger}\big]b_s^{\dagger a}(k^+,\bar{z})b_s^a(k^+,z)\big[G\hat{S}^\dagger-\hat{S}G\big]|v\rangle \\ 
&=&\frac{\alpha_s}{2\pi^2}\frac{1}{(2\pi)^2}\int e^{ik(z-\bar{z})}\frac{1}{p^+}\left(1+\xi^2\right)\frac{(v-\bar{z})\cdot(v-z)}{(v-\bar z)^2(v-z)^2}\langle b_s^{\dagger \bar{\sigma}}\big(\frac{k^+}{\xi},\bar u\big)b_s^{\sigma}\big(\frac{k^+}{\xi},u\big)\rangle\nonumber\\
&\times&\left\{\bigg[S_F^{\dagger}\big(\bar u\big)\tau^a\bigg]_{\bar{\sigma}\kappa}-S^{\dagger ba}(v)\bigg[\tau^{b}S^{\dagger}_F(\bar{z})\bigg]_{\bar{\sigma}\kappa}\right\} 
\left\{\bigg[\tau^aS_F\big(u\big)\bigg]_{\kappa \sigma}-S^{ac}(v)\bigg[S_F(z)\tau^{c}\bigg]_{\kappa \sigma}\right\}\nonumber \\ 
&+&\frac{\alpha_s}{2\pi^2}\frac{1}{(2\pi)^2}\int e^{ik(z-\bar{z})}\frac{1}{p^++k^+}\left(\xi^2+(1-\xi)^2\right)\frac{(v-\bar{z})_i(v-z)_j}{(v-\bar z)^2(v-z)^2}\langle a_{i}^{\dagger c}\big(\frac{k^+}{\xi},\bar u\big)a_j^{b}\big(\frac{k^+}{\xi},u\big)\rangle \nonumber\\
&\times & tr\bigg\{\bigg[S_F\big(u\big)\tau^{b}S_F^{\dagger}\big(u\big)-S_F(z)\tau^{b}S_F^{\dagger}(v)\bigg]
\bigg[S_F\big(\bar u\big)\tau^{c}S_F^{\dagger}\big(\bar u\big)-S_F(v)\tau^{c}S_F^{\dagger}(\bar{z})\bigg]\bigg\} \nonumber
\end{eqnarray}
The elastic contribution is
\begin{eqnarray}
\Bigg[\frac{dN_q}{d^2kd\eta}\Bigg]_{elastic}&=&\frac{\alpha_s}{4\pi^2}\frac{N_c^2-1}{N_c}\frac{1}{(2\pi)^2}\int_{x_F}^1\frac{d\xi}{\xi} e^{ik(z-\bar{z})}\frac{1+\xi^2}{1-\xi}\frac{(v-\bar{z})\cdot(v-z)}{(v-\bar z)^2(v-z)^2}\frac{1}{N_c}tr\big(S_F^{\dagger}(z)S_F(\bar{z})\big)\frac{k^+}{2\pi\xi}\langle b_s^{\dagger \sigma}\big(\frac{k^+}{\xi},\bar u\big)b_s^{\sigma}\big(\frac{k^+}{\xi},u\big)\rangle\nonumber \\
&+&\frac{\alpha_s}{8\pi^2}\frac{1}{(2\pi)^2}\int_{x_F}^1\frac{ d\xi}{\xi} e^{ik(z-\bar{z})}\left(\xi^2+(1-\xi)^2\right)\frac{(v-\bar{z})\cdot(v-z)}{(v-\bar z)^2(v-z)^2}\frac{1}{N_c}tr\big(S_F^{\dagger}(z)S_F(\bar{z})\big) \nonumber \\ &\times&\frac{k^+}{2\pi\xi}
\langle a_j^{\dagger \sigma}\big(\frac{k^+}{\xi},\bar u\big)a_j^{\sigma}\big(\frac{k^+}{\xi},u\big)\rangle\\
&\simeq&\frac{\alpha_s}{\pi}\frac{1}{(2\pi)^2}\int_{p^2<Q^2}\frac {dp}{p}\int_{x_F}^1\frac{d\xi}{\xi}x_F\Bigg[P_{q/q}(\xi)f_q(\frac{x_F}{\xi},p^2)+P_{q/g}(\xi)f_{g}(\frac{x_F}{\xi},p^2)\Bigg]N_F(k)\nonumber
\end{eqnarray}
with
\begin{equation}
P_{q/q}=\frac{N_c^2-1}{2N_c}\frac{1+\xi^2}{1-\xi}; \ \ \ \ \ P_{q/g}=\frac{1}{2}\left(\xi^2+(1-\xi)^2\right)
\end{equation}
and the inelastic term is 
\begin{eqnarray}
\Bigg[\frac{dN_q}{d^2kd\eta}\Bigg]_{inelastic}&=&\frac{\alpha_sN_c}{8\pi^2}\frac{1}{(2\pi)^2}\int_{x_F}^1\frac{d\xi}{\xi} e^{ik(z-\bar{z})}\frac{1+\xi^2}{1-\xi}\bigg[(1+\xi^2)-(1-\xi)^2\frac{1}{N_c^2}\bigg]\frac{[(v-\bar{z})\cdot(v-z)]^2}{(v-\bar{z})^2(v-z)^2}\frac{1}{N_c}tr\big(\partial_kS^{\dagger}_v\partial_kS_v\big)\nonumber \\ &\times&\frac{k^+}{2\pi\xi}
\langle b_s^{\dagger \sigma}\big(\frac{k^+}{\xi},\bar u\big)b_s^{\sigma}\big(\frac{k^+}{\xi},u\big)\rangle\nonumber \\
&+&\frac{\alpha_s}{16\pi^2}\frac{1}{(2\pi)^2}\int_{x_F}^1 \frac{d\xi}{\xi} e^{ik(z-\bar{z})}\left(\xi^2+(1-\xi)^2\right)\bigg[\big[(1-\xi)^2+\xi^2\big]-\frac{2\xi(1-\xi)}{N_c^2-1}\bigg]\frac{[(v-\bar{z})\cdot(v-z)]^2}{(v-\bar{z})^2(v-z)^2}\nonumber \\ 
&\times&\frac{1}{N_c}tr\big(\partial_kS^{\dagger}_v\partial_kS_v\big)
\frac{k^+}{2\pi\xi}
\langle a_j^{\dagger \sigma}\big(\frac{k^+}{\xi},\bar u\big)a_j^{\sigma}\big(\frac{k^+}{\xi},u\big)\rangle\\
&=&\frac{\alpha_s}{2\pi^2}\frac{1}{k^4}\int_{p^2<Q^2}\frac{d^2p}{(2\pi)^2} p^2 N_F(p)
x_F\int_{x_F}^1\frac{d\xi}{\xi}\nonumber\\
&\times&\Bigg[\frac{N_c^2}{ N_c^2-1}\bigg[1+\xi^2-\frac{(1-\xi)^2}{N_c^2}\bigg]P_{q/q}(\xi)f_q(\frac{x_F}{\xi}, Q)
+\frac{1}{2}\bigg[(1-\xi)^2+\xi^2-\frac{2\xi(1-\xi)}{N_c^2-1}\bigg]P_{q/g}(\xi)f_g(\frac{x_F}{\xi}, Q)\Bigg]
\nonumber
\end{eqnarray}

For antiquarks the calculation is identical.
The elastic contribution is
\begin{equation}
\Bigg[\frac{dN_{\bar q}}{d^2kd\eta}\Bigg]_{elastic}
\simeq\frac{\alpha_s}{\pi}\frac{1}{(2\pi)^2}\int_{p^2<Q^2}\frac {dp}{p}\int_{x_F}^1\frac{d\xi}{\xi}x_F\Bigg[P_{q/q}(\xi)f_{\bar q}(\frac{x_F}{\xi},p^2)+P_{q/g}(\xi)f_{g}(\frac{x_F}{\xi},p^2)\Bigg]N_F(k)
%
\end{equation}
and the inelastic term is 
\begin{eqnarray}
\Bigg[\frac{dN_{\bar q}}{d^2kd\eta}\Bigg]_{inelastic}
&=&\frac{\alpha_s}{2\pi^2}\frac{1}{k^4}\int_{p^2<Q^2}\frac{d^2p}{(2\pi)^2} p^2 N_F(p)
x_F\int_{x_F}^1\frac{d\xi}{\xi}\\
&\times&\Bigg[\frac{N_c^2}{ N_c^2-1}\bigg[1+\xi^2-\frac{(1-\xi)^2}{N_c^2}\bigg]P_{q/q}(\xi)f_{\bar q}(\frac{x_F}{\xi}, Q)
+\frac{1}{2}\bigg[(1-\xi)^2+\xi^2-\frac{2\xi(1-\xi)}{N_c^2-1}\bigg]P_{q/g}(\xi)f_g(\frac{x_F}{\xi}, Q)\Bigg]\nonumber
\end{eqnarray}
It is convenient to introduce the following functions, which we will dub "inelastic weights"
\begin{eqnarray}\label{weights}
w_{g/g}(\xi)&=&2\frac{N_c^2}{ N_c^2-1}(1-\xi+\xi^2)\\
w_{g/q}(\xi)&=&w_{g/\bar q}(\xi)=\frac{N_c^2}{ N_c^2-1}\bigg[1+(1-\xi)^2-\frac{\xi^2}{N_c^2}\bigg]\\
w_{q/q}(\xi)&=&w_{\bar q/\bar q}(\xi)=\frac{N_c^2}{ N_c^2-1}\bigg[1+\xi^2-\frac{(1-\xi)^2}{N_c^2}\bigg]\\
w_{q/g}(\xi)&=&w_{\bar q/g}(\xi)=\frac{1}{2}\bigg[(1-\xi)^2+\xi^2-\frac{2\xi(1-\xi)}{N_c^2-1}\bigg]
\end{eqnarray}
In terms of these functions one can write the 
inelastic contribution to production of $i$'th partonic flavor as
\begin{equation}
\Bigg[\frac{dN_{i}}{d^2kd\eta}\Bigg]_{inelastic}=\frac{\alpha_s}{ 2\pi^2}\frac{1}{k^4}\int_{p^2<Q^2}\frac{d^2p}{(2\pi)^2} p^2 N_F(p)x_F\int_{x_F}^1\frac{d\xi}{\xi}\Sigma_{j=q,\bar q, g}w_{i/j}(\xi)P_{i/j}(\xi)f_j(\frac{x_F}{\xi}, Q)
\end{equation}

Finally, including the effects of parton fragmentation, the result for hadron production is
\begin{eqnarray}\label{final}
\frac{dN_h}{d^2kd\eta}&=&\frac{1}{(2\pi)^2}\int_{x_F}^1 \frac{dz}{z^2} \Bigg[x_1f_g(x_1,Q^2)N_A(x_2,\frac{k}{z})D_{h/g}(z,Q)+\Sigma_qx_1f_q(x_1,Q^2)N_F(x_2,\frac{k}{z})D_{h/q}(z,Q)\Bigg]\nonumber\\
&+&\int_{x_F}^1 \frac{dz}{z^2}\frac{\alpha_s}{ 2\pi^2}\frac{z^4}{k^4}\int_{p^2<Q^2}\frac{d^2p}{(2\pi)^2} p^2 N_F(p,x_2)x_1\int_{x_1}^1\frac{d\xi}{\xi}\Sigma_{j=q,\bar q, g}w_{i/j}(\xi)P_{i/j}(\xi)f_j(\frac{x_1}{\xi}, Q)D_{h/q}(z,Q)
\end{eqnarray}
where the momentum fractions $x_1$ and $x_2$ are defined in eq.(\ref{xs}) and the inelastic weights $w_i$ in eq.(\ref{weights}).
This is our final result.

\section{Discussion}
In this paper we have derived the complete leading twist expression for inclusive hadron production in the hybrid formalism. We have shown that in addition to elastic scattering terms first derived in\cite{adrian}, there are also terms that correspond to inelastic scattering of the projectile partons on low momentum components of the target field. 
These terms are given by the second line in eq.(\ref{final}). We note that although the inelastic piece has an explicit factor of $\alpha_s$ while the elastic contribution does not, the two terms at high $k_T$ are in fact of the same order in $\alpha_s$. The reason is that at momenta $k\gg Q_s$ the dipole scattering amplitude $N_{A(F)}(k)$, which enters the elastic scattering term is itself of order $\alpha_s$, while the integral of the amplitude appearing in the inelastic term is of order unity.

The final states that correspond to the inelastic process are dihadron pairs where both hadrons are emitted at forward rapidity and have strong back to back correlation. Since both produced hadrons have large rapidity, such pairs with large transverse momentum are kinematically allowed only at large collision energy. Thus one might expect this contribution not to be of great importance in RHIC kinematics, however it may be sizable at LHC.

In this context we believe that including this contribution in calculation {\it a la} \cite{albacete} should produce faster approach of nuclear modification factor  $R_{pA}$ to unity at large transverse momenta. Here we wish to elaborate on possible role of saturation in the results of \cite{albacete}. 
As we have noted above, as long as the transverse momentum is above $Q_s$, saturation should mainly affect the inelastic production piece. This contribution involves the target distribution  $\int_{p^2<Q^2}\frac{d^2p}{(2\pi)^2} p^2 N_F(p,x_2)\propto f_{target}(Q,x_2)$ and is thus directly sensitive to saturation effects which suppress the contribution of small momenta $p< Q_s$ to the integral. 
The elastic production probability (first line of eq.(\ref{final})) depends only on $N(k)$ at large momentum. Naively one expects that this part is unaffected by saturation in the evolution. This does not necessarily mean that the $R_{pA}$ calculated using only this contribution (as done in \cite{albacete}) should be equal to unity, but rather that any departure from unity is the effect of a non scaling initial condition. This should be true if the transverse momentum of the measured particle is
 in the so called "`geometric scaling"' window, where the anomalous dimension is finite, since geometric scaling is not a result of saturation physics but rather of the linear BFKL evolution of the gluon density\cite{geometric}.  To be a little more precise, recall that solution of the BFKL equation above the saturation scale has the form
 $\phi_{BFKL}(k,Y)\propto [Q_s(Y)/k]^{2-2\gamma}$ where $\gamma$ is the anomalous dimension. The anomalous dimension is a slowly varying function of transverse momentum. It is almost constant in a wide window of momenta above $Q_s$, but nevertheless vanishes asymptotically as $k\rightarrow\infty$. It also weakly depends on rapidity $Y$. The saturation momentum $Q_s$ is defined withing the BFKL solution {\it per se} as the momentum at which the scattering amplitude is of order one. Within leading order BFKL solution $Q_s(Y)=Q_0\exp\{\lambda Y\}$ where $Q_0$ is the soft nonperturbative scale which characterizes the initial condition $\phi_0(k,Y=0)$. In the case of calculations of \cite{albacete} this would be the initial saturation scales, $Q_{0p}$ for the proton target and $Q_{0A}$ for the nuclear target. The nuclear modification factor $R_{pA}$ within a BFKL calculation would then be 
\begin{equation}
R_{pA}(Y)=\frac{1}{N_{coll}}\Bigg[\frac{Q_{sA}(Y)}{Q_{sp}(Y)}\Bigg]^{2-2\gamma(Y)}=\Bigg[\frac{Q_{0p}}{Q_{0A}}\Bigg]^{2\gamma(Y)}
\end{equation}
with the identification $N_{coll}=Q^2_{0A}/Q^2_{0p}$. Within the running coupling calculation the saturation scale is not a simple exponential of rapidity and thus the explicit expression for the nuclear modification factor and the rapidity dependence is somewhat different. It nevertheless remains the case that as long as the initial conditions for proton and nucleus do not simply scale with $A^{1/3}$ at all momenta, $\phi_p(k,Y=0)\ne A^{1/3}\phi_A(k,Y=0)$, the linear BFKL evolution produces a nuclear modification factor different from unity and slowly varying with rapidity. It is an interesting question whether the numerical results of \cite{albacete} are consistent with BFKL, or whether saturation effects in the evolution nevertheless give a significant contribution to $R_{pA}$.

Finally we note that the final states that contribute to the inelastic production are precisely the states which give the bulk of the contribution to the dihadron correlation function considered in \cite{albacete1}. The calculation of \cite{albacete1} does not address the estimate of large uncorrelated background of produced particles. The small "`signal to background"' ratio is indeed a very pronounced feature of the data\cite{corr},\cite{mark}. In this respect it would be interesting to calculate both in a unified framework discussed here. We note that our earlier discussion suggests that the saturation has two distinct effects on the correlated dihadron production. First, as discussed in \cite{baier}, \cite{marquet} and \cite{albacete}, the back-to-back correlation is weakened due to independent momentum transfer from the target to each one of the produced hadrons. This does not reduce the number of hadrons produced at forward rapidity, but reduces the correlation between the direction of their transverse momenta. Another distinct effect is that the dihadron production probability is suppressed by the effect of saturation on  $\int_{p^2<Q^2}\frac{d^2p}{(2\pi)^2} p^2 N_F(p,x_2)$, thus reducing the ratio of the correlated signal to the total number of produced particles.

\section*{Acknowledgments}
We thank Javier Albacete and Cyrille Marquet for useful correspondence. This work was supported by the DOE grant DE-FG02-92ER40716.



\begin{thebibliography}{99}
\bibitem{footnote1} There is also a large number of works applying the saturation ideas to heavy ion collisions both at RHIC and LHC. The analysis of these processes however is obscured by very complicated final state interactions and ensuing collective effects, and thus at present are not at the stage of reliable quantitative predictions/descriptions.
\end{thebibliography}
\end{document}